# OBSERVATIONAL ASPECTS OF WAVE ACCELERATION IN OPEN MAGNETIC REGIONS

## Steven R. Cranmer

Harvard-Smithsonian Center for Astrophysics, Cambridge, MA 02138, USA
*Email:* scranmer@cfa.harvard.edu

## ABSTRACT

This paper reviews the latest observational evidence for the existence of propagating waves in the open magnetic flux tubes of the solar corona. *SOHO* measurements have put tentative limits on the fluxes of various types of magnetohydrodynamic (MHD) waves in the acceleration region of the solar wind. Also, continually improving measurements of fluctuations at larger distances (i.e., *in situ* detection and radio scintillation) continue to provide significant constraints on the dominant types of plasma oscillation throughout the corona and wind. The dissipation of MHD fluctuations of one kind, probably involving anisotropic turbulent cascade, is believed to dominate the heating of the extended corona. Spectroscopic observations from the UVCS instrument on *SOHO* have helped to narrow the field of possibilities for the precise modes, generation mechanisms, and damping channels. This presentation will also review some of the collisionless, kinetic aspects of wave heating and acceleration that are tied closely to the observational constraints.

Key words: coronal holes; MHD waves; solar corona; solar wind; plasma physics; turbulence; UV spectroscopy.

## 1. INTRODUCTION

The physical processes responsible for transporting the mechanical energy of sub-photospheric convective motions into the corona and converting this energy into heat remain unknown after more than a half-century of study. Different processes for heating the corona probably govern closed loops, bright points, and the large-scale open field lines that feed the solar wind (e.g., Narain and Ulmschneider, 1990, 1996; Priest et al. 2000). There is also a growing realization that the "base" of the corona ($r \lesssim 1.5\,R_\odot$) is probably heated by different mechanisms than those that are dominant at larger distances from the Sun. Theoretical models of this latter region, often called the "extended corona," typically involve the transfer of energy from *propagating fluctuations* (i.e., waves, shocks, or turbulent eddies) to the particles. This general consensus has arisen because the ultimate source of energy must be the Sun itself, and thus the energy must somehow propagate out to the distances where the extended heating occurs. Obtaining empirical evidence for these propagating fluctuations has been a priority in solar physics, and this paper summarizes their properties from direct and indirect measurements.

Alfvén (1947) performed one of the first studies of the role of magnetohydrodynamic (MHD) waves in the heating of the extended corona. Interestingly, one of his aims was to counter the popular (but unattributed) idea that the corona could be heated by meteors falling toward the Sun and being stopped by friction. In the intervening decades much progress has been made in the study of waves in open magnetic field regions. As of September 2003, a search of NASA's Astrophysics Data System (ADS) yielded approximately 4600 abstracts that contain both "waves" and "solar wind," one third of which are in the last 10 years. This paper can cite only a small fraction of this work, but many other comprehensive reviews have been written (see, e.g., Hundhausen 1972; Hollweg 1974; Leer et al. 1982; Parker 1991; Velli 1994, 1999; Tu and Marsch 1995; Goldstein et al. 1995; Feldman and Marsch 1997; Marsch 1999; Roberts 2000; Hollweg and Isenberg 2002; Cranmer 2002a).

This paper is organized as follows. Direct measurements of wavelike oscillations are reviewed in § 2, and indirect measurements (i.e., inferences made about wave properties from specific kinds of ion heating) are reviewed in § 3. Brief summaries of the challenges remaining to theorists and observers are given in §§ 4–5.

## 2. DIRECT MEASUREMENTS

Linear, wavelike oscillations in plasma reveal themselves by presenting fluctuations in velocity, density, pressure, and magnetic field strength. Waves in ideal MHD (applicable to the largest scales in the low-beta solar corona) sort themselves into three types: transverse Alfvén waves, slow-mode magnetosonic waves (which behave like simple acoustic waves when propagating along the field), and fast-mode magnetosonic waves. When wave spatial scales grow smaller than typical plasma inertial lengths or Larmor radii, the number of possible *kinetic* wave modes grows rapidly (see Oraevsky 1983; Stix 1992; Brambilla 1998). *In situ* measurements of wave properties are beginning to detect small-scale kinetic effects (e.g., Leamon et al. 1998), but remote-sensing measurements generally cannot make clear distinctions between ideal MHD and kinetic wave properties. Also, the quantities measured remotely are often most sensitive to the wave modes that carry the most energy, which are probably dominated by the longest wavelengths. Thus, the direct measurements described below are typically interpreted as ideal MHD fluctuations.

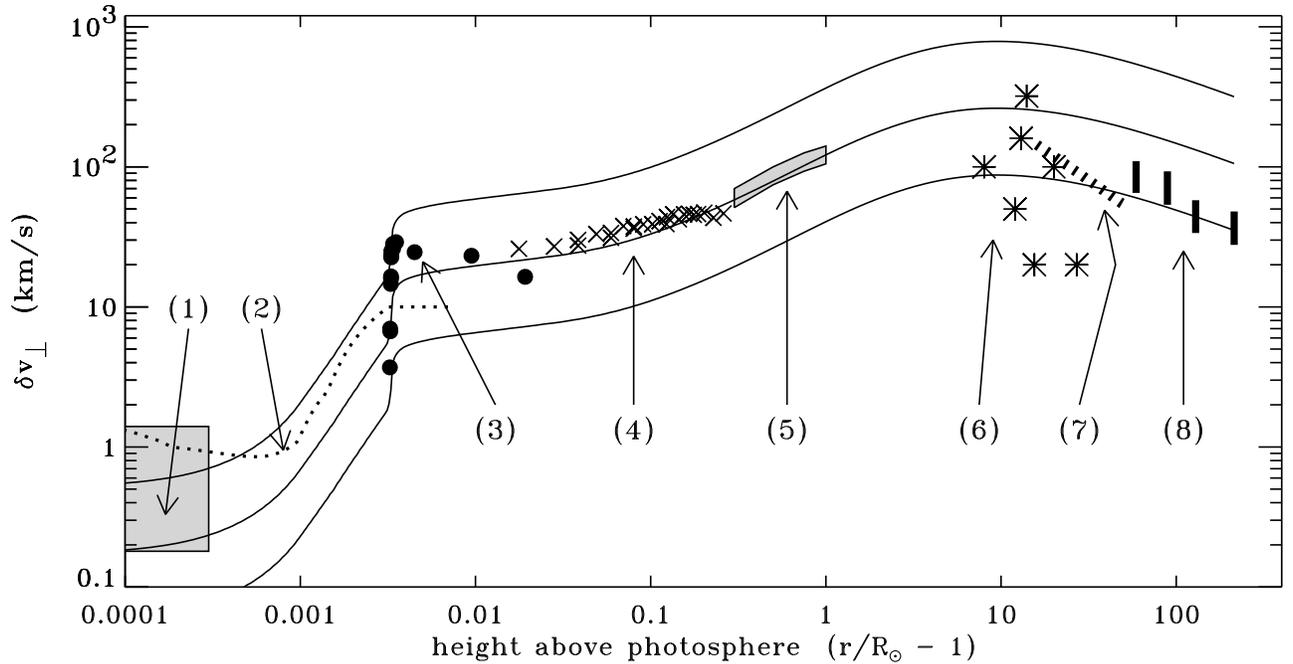

*Figure 1. Assembled plot of empirically derived **velocity fluctuation amplitudes** in the corona and fast solar wind. Numbered sets of measurements are discussed further in the text. Parallel solid curves denote how the transverse velocity amplitudes of linear Alfvén waves would behave as a function of height (with three different normalizations) assuming WKB wave action conservation.*

Fig. 1 displays eight sets of observational determinations of transverse velocity fluctuations ($\delta v_\perp$) at a range of heights in the solar corona (weighted toward coronal holes and the high-speed wind). These motions are generally believed to indicate the amplitudes of incompressible Alfvén waves. It should be emphasized that the juxtaposition of different $\delta v_\perp$ values in Fig. 1 is, to some degree, uncritical of the respective systematic uncertainties of the various types of measurement and blind to the observed ranges of frequency. Differences in the precise definitions of the velocity amplitude also imply that the values are uncertain to within multiplicative factors of order $\sqrt{2}$.

From low to high heliocentric distances, the numbered sets of observations are described in more detail below.

1. Transverse motions in the photosphere are often probed by high-resolution time-series observations of G-band bright points in intergranular lanes. The mean transverse velocities inferred from these observations can be as large as $1.4 \text{ km s}^{-1}$ (e.g., Muller et al. 1994; Nisenson et al. 2003), but only some fraction of that will likely propagate upwards on open field lines; van Ballegooijen et al. (1998) estimated this amplitude to be of order $0.2 \text{ km s}^{-1}$. This range of speeds is plotted in Fig. 1.

2. Empirically driven energy-balance models of the photosphere, chromosphere, and transition region must include an *ad hoc* "microturbulence" in order to match the widths of lines formed at a range of heights. The values of $v_{\text{turb}}$ plotted here come from a recent quiet-Sun model (E. Avrett 2003, personal communication; see also Fontenla et al. 1993). The unresolved motions are assumed to have a Gaussian distribution of random velocities.

3. SUMER/*SOHO* has observed "nonthermal" broadening velocities of various ions by assuming that spectral lines are formed mainly in plasma with an electron temperature $T_e$ equal to the peak temperature of the equilibrium ionization balance for each ion. Broader lines imply an additional unresolved velocity attributable to wave motions along the line of sight. In the quiet Sun, Chae et al. (1998) measured 17 individual nonthermal velocities as a function of $T_e$, and in Fig. 1 these values are mapped to height using the empirical temperature distribution of the above-cited energy-balance model.

4. Similar SUMER measurements made above the limb (Banerjee et al. 1998) have been used to infer the radial dependence of $\delta v_\perp$ at larger heights.

5. UVCS/*SOHO* spectroscopy at still larger heights has been used to constrain the velocity amplitude of unresolved wave motions by insisting that the proton temperature is not larger than the temperature of $Mg^{9+}$ ions (Esser et al. 1999). Wave action conservation (see below) was used to further constrain the radial dependence of $\delta v_\perp$ over the heights observed by UVCS.

6. Interplanetary scintillation (IPS) observations of radio signals passing through the corona allow some properties of plasma irregularities to be determined. One way of detecting random fluctuations in the bulk solar wind is by measuring departures from a "frozen-in" diffraction pattern measured by different sets of radio receivers. The stars plotted in Fig. 1 denote an early attempt (Armstrong and Woo 1981) to separate the bulk solar wind flow speed from the random wavelike component, interpreted as $\delta v_\parallel$.

7. A more recent IPS determination of (specifically transverse) velocity fluctuations in the fast wind was performed by Canals and Breen (2000) using the EISCAT facility, with $\delta v_\perp \propto r^{-0.9}$ between 17 and 50 $R_\odot$.

8. The *Helios* 1 and 2 probes uniquely measured the *in situ* plasma properties between Mercury and the Earth, and they measured MHD fluctuations spanning a wide range of time scales. The integrated magnetic fluctuation amplitude $\delta B$, discussed by Tu (1987), has been converted into a velocity amplitude by assuming the ideal MHD Wálen condition $(\delta v_\perp / V_A) = (\delta B / B)$ to hold in fast streams (e.g., Tu and Marsch 1995; Goldstein et al. 1995).

The parallel solid curves in Fig. 1 show how the transverse velocity amplitude of Alfvén waves should behave as a function of height under the assumption of linear wave action conservation; i.e.,

$$\delta v_\perp \propto \frac{\sqrt{u_\parallel V_A}}{u_\parallel + V_A} \qquad (1)$$

where $u_\parallel$ is the bulk outflow speed of the wind and $V_A$ is the Alfvén speed (e.g., Jacques 1977). The adopted values of $u_\parallel(r)$ and $V_A(r)$ come from the empirical magnetic field model of Banaszkiewicz et al. (1998), a mean whitelight number density as given by, e.g., Guhathakurta and Holzer (1994) and Fisher and Guhathakurta (1995), and mass flux conservation. Note that, under wave action conservation, the relationship between $\delta v_\perp$ and the mean density $\rho_0$ varies significantly from the hydrostatic atmosphere to the distant solar wind. At low heights, $\delta v_\perp \propto \rho_0^{-1/4}$, but at large distances from the Sun, $\delta v_\perp \propto \rho_0^{+1/4}$. (This is valid for homogeneous flux tubes that have constant mass flux $[\rho_0 u_\parallel A]$, but this may not hold true for inhomogeneous flux bundles; Moran 2001).

Inspection of Fig. 1 shows that most of the direct $\delta v_\perp$ measurements seem to roughly obey linear wave action conservation, but it seems necessary for some damping to occur above $\sim 10$ $R_\odot$. The local "peak" in the Chae et al. (1998) SUMER data (at $T_e \approx 3 \times 10^5$ K, mapped here to a height of $\sim 0.003$ $R_\odot$) seems not to obey wave action conservation—possibly implying substantial *wave reflection* at this height—but this could be an artifact of my simplistic mapping from "formation temperature" to height.

Fig. 2 displays six sets of observational determinations of compressive density fluctuations, expressed dimensionlessly as $\delta\rho$ divided by the background density $\rho_0$. Many of the same caveats concerning Fig. 1 apply to this assemblage of observations as well. The numbered sets of measurements are described in more detail below.

1. There are many observations of compressive fluctuations in the chromosphere, measured either as intensity fluctuations $\delta I$ or as "longitudinal" Doppler shifts $\delta v_\parallel$. The representative data point plotted in Fig. 2 comes from a time series of Fe I Doppler shifts (Lites et al. 1993; Theurer et al. 1997), mapped into density fluctuations by assuming equipartition between kinetic and thermal motions (i.e., $\delta v_\parallel / c_s \approx \delta\rho / \rho_0$, where $c_s$ is the acoustic sound speed).

2. The transition region is dominated by large-amplitude intensity fluctuations. Limits on wavelike amplitudes in coronal holes and the quiet Sun are plotted here (see Pérez et al. 1999; Innes 2001).

3. Early in the *SOHO* mission, the EIT instrument observed upward-propagating intensity oscillations in bright polar plumes, which have been interpreted as slow magnetosonic waves (DeForest and Gurman 1998; Ofman et al. 1999).

4. At larger heights, measurements of the visible polarization brightness (pB) with the UVCS White Light Channel were also used to infer the presence of density oscillations (Ofman et al. 1997).

5. Radio IPS measurements are sensitive to density fluctuations over a wide range of spatial scales. Spangler (2002) presented integrated values of $\delta\rho / \rho_0$ from VLBI measurements, and compared them to predictions based on specific MHD modes.

6. The *in situ* density fluctuation spectra between 0.3 and 1 AU show a large intrinsic variability, with no clear radial trend discernible (e.g., Tu and Marsch 1994).

The solid and dashed curves in Fig. 2 show the expected radial dependence of $\delta\rho / \rho_0$ for wave action conservation (of acoustic or parallel slow-mode waves), and the measurements above a height of $\sim 0.03$ $R_\odot$ seem roughly consistent with the solid curves. However, the data points in the chromosphere and transition region imply much larger amplitudes which rapidly reach nonlinear saturation ($\delta\rho \approx \rho_0$). This is consistent with long-standing predictions that acoustic waves must saturate, steepen into shocks, and damp before they can penetrate far into the corona (e.g., Athay and White 1978; Cuntz and Suess 2001). It is still not known whether the small fraction of compressible power "left over" at larger heights is simply a remnant of the waves that were damped, or a secondary local generation (from, e.g., nonlinear evolution of the dominant population of Alfvén waves).

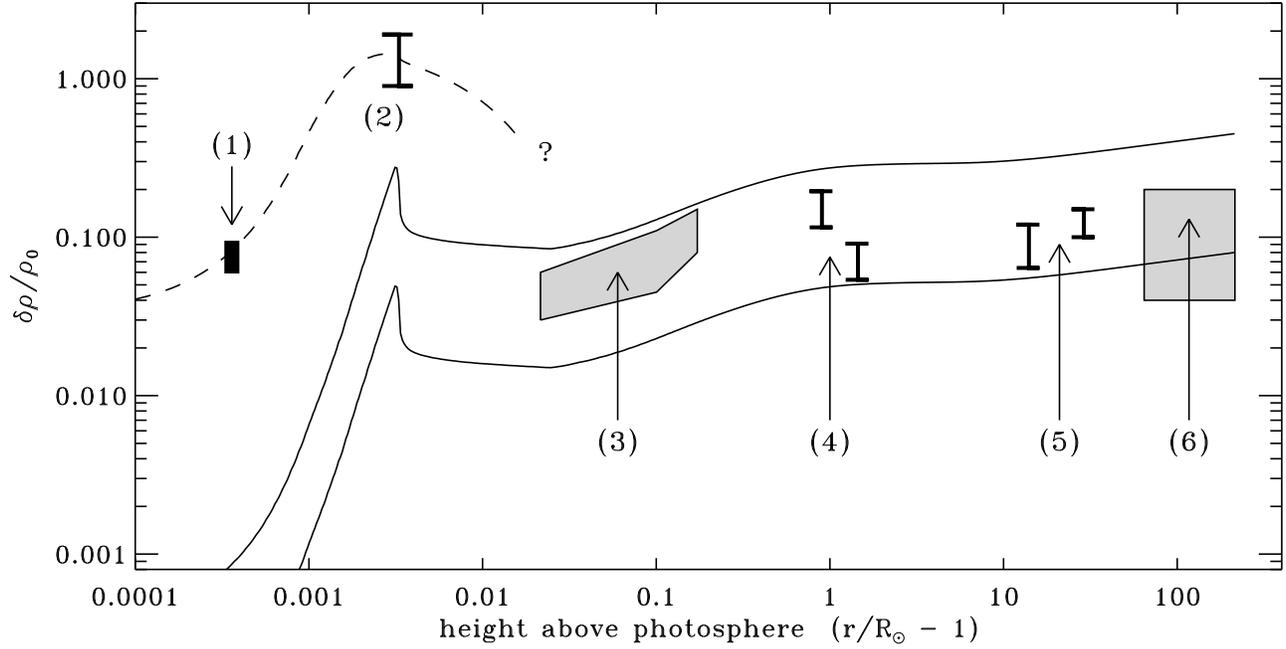

*Figure 2. Assembled plot of empirically derived **density fluctuation amplitudes** in the corona and fast solar wind. Numbered sets of measurements are discussed further in the text. Parallel solid curves denote how the fractional density amplitudes of linear acoustic waves would behave as a function of height (with two different normalizations) assuming WKB wave action conservation, and the dashed curve indicates a stronger fractional amplitude that is saturated around $\delta\rho \approx \rho_0$ and damped with an exponential dissipation length of $0.006\ R_\odot$.*

## 3. INDIRECT MEASUREMENTS

If waves in the extended corona just expand passively outward on open field lines without interacting with the background plasma, they would not be of much interest. The exact manner in which the plasma is heated and accelerated by wave-particle interactions, then, is useful as a means of determining the wave modes that are generated and damped.

It is fortuitous that the density in the extended corona is as *low* as it is, because the relative lack of Coulomb collisions allows each particle species (electrons, protons, and heavy ions) to react to the waves in its own unique fashion. SUMER measurements have shown that ion temperatures exceed electron temperatures at very low heights (Seely et al. 1997; Tu et al. 1998). Coronal electrons may have non-Maxwellian velocity distributions reminiscent of the "core" and "halo" measured *in situ* (Pinfield et al. 1999; Esser and Edgar 2000; Chen et al. 2003).

At the last solar minimum, UVCS provided the first measurements of preferential ion heating, ion temperature anisotropies, and differential outflow speeds in the acceleration region of the wind (Kohl et al. 1997, 1998, 1999). In coronal holes, UVCS measured $O^{5+}$ perpendicular temperatures exceeding 100 million K at heights above 2 $R_\odot$ (see Fig. 3). The Doppler dimming and pumping of the individual lines of the O VI doublet allowed the anisotropy ratio $T_\perp/T_\parallel$ of the $O^{5+}$ ions to be constrained to values of at least 10, and possibly as large as 100. Temperatures for both $O^{5+}$ and $Mg^{9+}$ are significantly greater than mass-proportional when compared to hydrogen, and outflow speeds for $O^{5+}$ may exceed those of hydrogen by as much as a factor of two (see also Li et al. 1998; Cranmer et al. 1999b; Giordano et al. 2000; Zangrilli et al. 2002).

Over the past 7 years of the *SOHO* mission, UVCS has observed a wide range of plasma conditions in coronal holes and streamers. Fig. 3 illustrates the range of observed $T_\perp$ values by means of the Doppler broadening of the O VI 1032, 1037 Å emission line doublet. (Strictly, this line broadening gives the projection of the ion velocity distribution along the line of sight; for a nearly radial magnetic field, this is most sensitive to $T_\perp$.) Definite ion anisotropies ($T_\perp > T_\parallel$) have been deduced for high-latitude coronal holes at solar maximum (Miralles et al. 2001b, 2002) and at large heights in equatorial streamers at solar minimum (Frazin et al. 2003).

Velocity distributions measured *in situ* have qualitatively similar properties as those measured by UVCS. In the high-speed solar wind, *Helios* 1 and 2 measured proton temperature anisotropies with $T_\perp > T_\parallel$ (Marsch et al. 1982a). Most, though not all, ion species appear to flow faster than the protons, and this velocity difference decreases with increasing radius and decreasing proton flow velocity (e.g., Hefti et al. 1998; Reisenfeld et al. 2001). The temperatures of heavy ions are significantly larger than proton and electron core temperatures. In the highest-speed wind, ion temperatures exceed simple mass proportionality (i.e., heavier ions have larger most-probable speeds), with $(T_{ion}/T_p) > (m_{ion}/m_p)$, for $m_{ion} > m_p$. Fig. 4 illustrates the wind-speed dependence of ion-to-proton temperature ratios using *Wind* data at 1 AU (Collier et al. 1996).

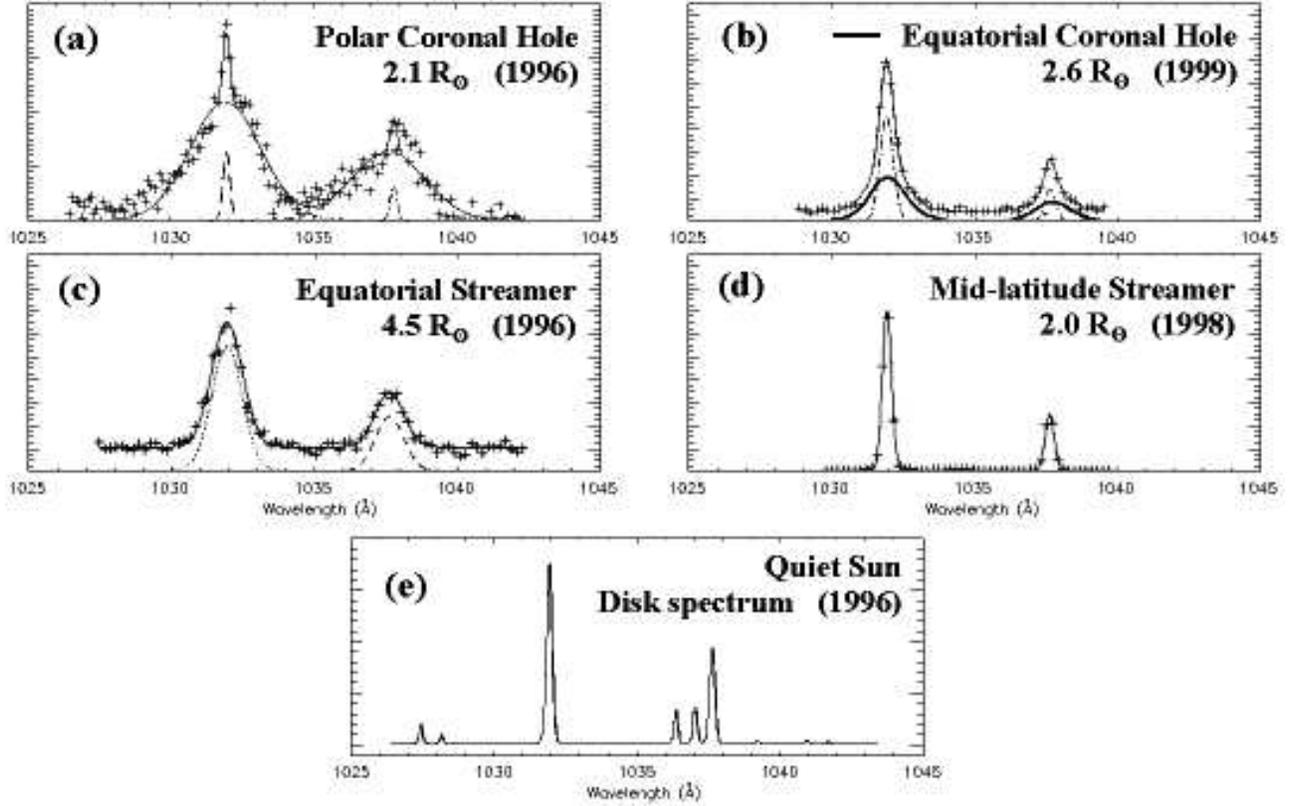

*Figure 3.* **(a)–(d)** *UVCS/SOHO observations of the O VI 1032, 1037 Å doublet in four types of coronal structure (see, e.g., Kohl et al. 1997; Frazin et al. 1999; Miralles et al. 2001a).* **(e)** *SUMER/SOHO observations of the quiet solar disk in the same range of the spectrum (Warren et al. 1997).*

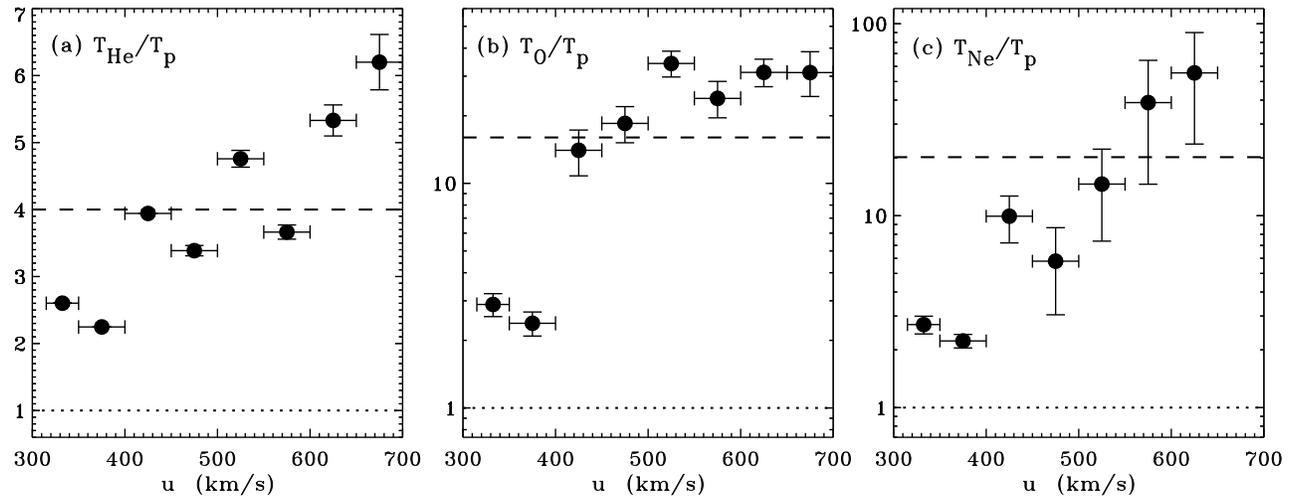

*Figure 4. Ion temperature ratios at 1 AU for varying solar wind speed (Collier et al. 1996). The proton temperatures were computed from the empirical relation $T_p = -0.240u^2 + 836u - 213000$, where $T_p$ is in K and u is in km s$^{-1}$. This is a fit to data presented by Ogilvie et al. (1980), for low speeds, and Goldstein et al. (1996), for high speeds. Dotted lines denote equal ion and proton temperatures, and dashed lines denote mass-proportional temperatures. Small differences between the proton and helium speeds (the latter used by Collier et al. 1996) are neglected in this plot.*

Note that when $T_{\rm ion}/T_p$ exceeds the mass ratio $m_{\rm ion}/m_p$, it is *impossible* to interpret the ion temperature as a combination of thermal equilibrium and a species-independent "nonthermal speed." The remote-sensing and *in situ* data have thus been widely interpreted as a truly "preferential" heating of heavy ions. However, Moran (2002) brought up the interesting alternative (for the SUMER off-limb data only) that there could be nonthermal broadening via certain types of dispersive waves that have a charge- and mass-dependent $\delta v_\perp$. This suggestion, though, assumes that the bulk of the wave power resides in large-frequency or small-scale kinetic modes.

In summary, the preponderance of evidence in both the extended corona and in interplanetary space points to the following properties of ions *in the fast solar wind*:

$$\left\{ \begin{array}{rcl} T_{\rm ion} & \gg & T_p > T_e \\[2mm] (T_{\rm ion}/T_p) & > & (m_{\rm ion}/m_p) \\[2mm] T_\perp & > & T_\| \\[2mm] u_{\rm ion} & > & u_p \end{array} \right\} \qquad (2)$$

Traditionally, in solar wind studies this collection of ion properties has been associated with the collisionless damping of *ion cyclotron resonant waves*—i.e., Alfvén waves with frequencies $\omega$ approaching the Larmor frequencies of the ions $\Omega_{\rm ion}$ (e.g., Toichi 1971; Harvey 1975; Abraham-Shrauner and Feldman 1977; Hollweg and Turner 1978; Marsch et al. 1982b; Isenberg and Hollweg 1983; Hollweg 1986; Tu 1987, 1988; Axford & McKenzie 1992). The *SOHO* observations discussed above have given rise to a resurgence of interest in ion cyclotron waves as a potentially important mechanism in the acceleration region of the fast wind (e.g., McKenzie et al. 1995; Tu and Marsch 1997, 2001, 2002; Fletcher and Huber 1997; Hollweg 1999; Li 1999, 2003; Cranmer et al. 1997, 1999a; Cranmer 2000, 2001, 2002a; Galinsky and Shevchenko 2000; Ofman et al. 2001, 2002; Vocks and Marsch 2001, 2002; Gary et al. 2003). There remains some controversy over the issue of whether ion cyclotron waves generated solely at the coronal base can heat the extended corona, or if more gradual, extended generation of these waves is needed (see Hollweg and Isenberg 2002, for a detailed summary).

Is the ion cyclotron resonance the only viable possibility? There have been several other mechanisms suggested for producing the above ion properties. If there is substantial power in obliquely propagating fast-mode waves, their collisionless damping may contribute to ion and proton heating (e.g., Li and Habbal 2001; Hollweg and Markovskii 2002). MHD waves propagating at large angles to the background magnetic field can steepen into shocks under certain conditions, and numerical simulations have produced a rich variety of steepening phenomena that produce power at high-frequency harmonics of an input spectrum (e.g., Spangler 1997). Certain types of collisionless shocks may also accelerate positive ions in the direction perpendicular to the magnetic field (Lee and Wu 2000).

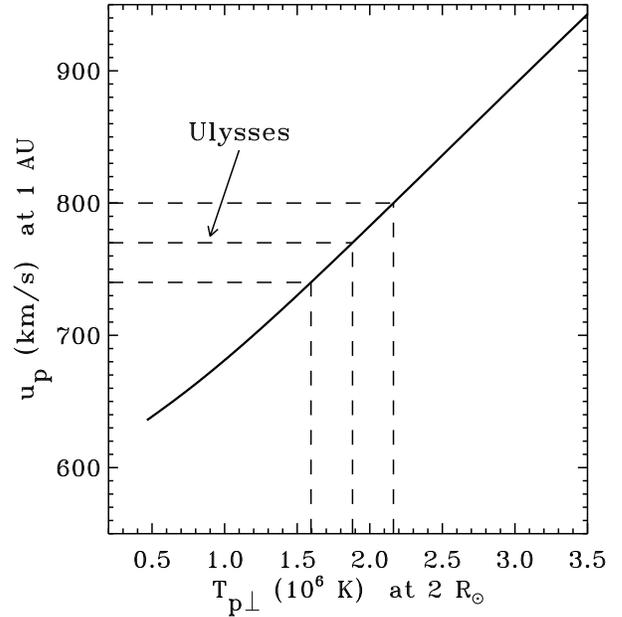

*Figure 5. Modeled wind outflow speed at 1 AU as a function of the perpendicular proton temperature at 2 $R_\odot$ (solid line). The mean and $\pm 1\sigma$ variations observed by Ulysses at solar minimum are plotted for comparison (dashed lines).*

In addition to the above proposed processes, there has been some very recent work on the complex interconnectedness between highly oblique kinetic Alfvén waves (KAW), perpendicular ion heating, and parallel electron beams (Markovskii and Hollweg 2002; Voitenko et al. 2003; Cranmer and van Ballegooijen 2003). Spacecraft measurements in the Earth's ionosphere and magnetosphere (e.g., Ergun et al. 1998, 1999) have provided clues to how ions can be heated perpendicularly in the presence of such phenomena (see also Matthaeus et al. 2003a, 2003b).

Finally, one additional "indirect" method of assessing the presence and impact of waves is by including the phenomenon of *wave pressure* in models of solar wind acceleration. Even the totality of the ion heating discussed above may not be enough to accelerate the fast solar wind to speeds exceeding 700–800 km s$^{-1}$ at 1 AU. In a radially inhomogeneous plasma such as the solar wind, the dissipationless propagation of MHD waves does work on the mean fluid (e.g., Belcher 1971; Jacques 1977) and provides an added outward acceleration. In order to make an empirically driven estimate of the importance of wave pressure, a series of solar wind models were computed by integrating the radial momentum equation (i.e., a generalized Parker critical point equation) assuming a parameterized radial dependence for $T_e$, $T_{p\|}$, and $T_{p\perp}$ and an imposed flux-tube area from the Banaszkiewicz et al. (1998) magnetic geometry (see Fig. 7 of Cranmer 2002a for details). A wave pressure term consistent with the middle curve of Fig. 1 was applied to these models; at $r = 2\,R_\odot$ this model has $\delta v_\perp = 120$ km s$^{-1}$. (A small degree of "saturation" was imposed above $\sim 10\,R_\odot$ to ensure that the ratio $\delta B/B$ never exceeded unity.) By varying the peak value of $T_{p\perp}$ in the corona, a continuous range of outflow speeds at 1 AU were produced.

Fig. 5 illustrates the results of these wave pressure models by plotting the outflow speed at 1 AU as a function of the perpendicular proton temperature at $r = 2\,R_\odot$. The range of *in situ* data from *Ulysses* are also shown (Goldstein et al. 1996), and the mean speed of $\sim$770 km s$^{-1}$ corresponds to $T_{p\perp} \approx 1.9 \times 10^6$ K at $2\,R_\odot$. The H I Ly$\alpha$ line width as observed by UVCS would be a convolution of the thermal and transverse wave speeds,

$$V_{1/e} = \left( \frac{2kT_{p\perp}}{m_p} + \delta v_\perp^2 \right)^{1/2} \approx 210\,\text{km s}^{-1}$$

$$(3)$$

which compares favorably with the empirically constrained value at that height of $203 \pm 11$ km s$^{-1}$ (Cranmer et al. 1999b). Despite this agreement, though, this result is far from conclusive evidence that we understand how the observed line profile is separated into thermal and $\delta v_\perp$ components. The assumed radial forms for the temperatures in this model were chosen in an observationally motivated, but still *ad hoc* manner. No effects from Coulomb energy exchange or heat conductivity were included (see, e.g., Li 1999, 2003). More work needs to be done, but the above result is at least suggestive that we are on the right track.

## 4. CURRENT CHALLENGES: COMPLEXITY AND TURBULENCE

Space plasmas always seem to be more complex when observed at high resolution than at low resolution. The direct and indirect measurements summarized above have helped constrain theoretical models, but they have raised many questions as well. Fig. 6a illustrates the multi-scale complexity that is suggested by the sum total of the observations at this time. It seems likely that the Sun launches a significant population of low-frequency ($f \lesssim 10^{-2}$ Hz) MHD waves, although the relative fluxes of fast, slow, and Alfvén waves are not yet known with certainty (see, though, Bogdan et al. 2002). At larger heights, these waves are somehow transformed into forms that lead to efficient heating and acceleration of the particles in the solar wind.

Fluctuations in the extended corona and solar wind are most likely highly turbulent, and indeed a nonlinear turbulent cascade has been suggested for some time to be able to transform low-frequency Alfvén waves into high-frequency ion cyclotron waves. However, both numerical simulations of MHD turbulence and analytic descriptions indicate that this cascade occurs most rapidly for transverse (high-$k_\perp$) fluctuations and hardly at all for fluctuations propagating along the field (high-$k_\parallel$; see, e.g., Shebalin et al. 1983; Matthaeus et al. 1996; Goldreich and Sridhar 1997; Bhattacharjee and Ng 2001; Cho et al. 2002). Alfvénic fluctuations having large $k_\perp$ and small $k_\parallel$ do not have high frequencies approaching the cyclotron resonances. Cranmer and van Ballegooijen (2003) investigated the possible kinetic consequences of this kind of anisotropic MHD turbulence, and modeled the cascade as a linear combination of advection and diffusion in $k_\perp$ space, characterized by dimensionless strengths $\beta$ and $\gamma$, respectively. Fits to the $k_\parallel$ dependence of MHD turbulence simulations seem to imply an exponential decline of power with increasing $k_\parallel$, which in the phenomenology of Cranmer and van Ballegooijen (2003) imply $\beta \gg \gamma$ (see Fig. 6b). Small values of $\gamma$, relative to $\beta$, imply negligible back-diffusion from large $k_\perp$ to the high-frequency (ion cyclotron resonant) region of $k$-space. However, the earlier statistical analysis of van Ballegooijen (1986) found that $\beta \approx \gamma$ is a reasonable assumption to make for field lines executing "random walks" at the photosphere. Furthermore, Cranmer and van Ballegooijen (2003) found that if $\beta/\gamma \lesssim 0.25$, there could indeed be enough wave energy at the ion cyclotron frequencies to heat protons in agreement with the observations (see Fig. 6c).

## 5. CURRENT CHALLENGES: OBSERVATIONS

Improvements in the observations are needed to make further progress in understanding the role of waves in open magnetic field regions in the corona and solar wind. (Only a subset of interesting future prospects are listed here.) The ultimate source of transverse wave motions in the corona may be in the motions of intergranular bright points in the photosphere, and G-band observations require extremely low image jitter (i.e., uncertainties less than $\sim$25 km) in order to measure velocity and vorticity power spectra. Space-based remote-sensing diagnostics of how waves heat and accelerate ions have the capability to be greatly improved, and next-generation coronagraph spectrometers are being designed with the capability to sample the velocity distributions of dozens of ions in the acceleration region of the fast wind in coronal holes. (For specific diagnostic predictions, see Cranmer 2002b.) Such instruments could also detect subtle departures from Gaussian line shapes that signal the presence of specific non-Maxwellian distributions—and thus specific wave-particle interactions (e.g., Cranmer 1998, 2001). New radio sounding techniques such as the measurement of scintillations in the circularly polarized Stokes parameters (e.g., Macquart and Melrose 2000) may be fruitful in extracting more information about MHD turbulence in the corona and solar wind.

## ACKNOWLEDGMENTS

The author would like to thank Aad van Ballegooijen, John Kohl, Gene Avrett, and Steve Spangler for many valuable discussions. The *SOHO* observations would not be possible without the tireless efforts of the instrument teams and operations personnel at the Experimenters' Operations Facility (EOF) at Goddard. Special thanks also go to all concerned with the rigorous calibration of the *SOHO* instruments (see, e.g., Pauluhn et al. 2002). This work is supported by the National Aeronautics and Space Administration under grants NAG5-12865 and NAG5-11913 to the Smithsonian Astrophysical Observatory, by Agenzia Spaziale Italiana, and by the Swiss contribution to the ESA PRODEX program.

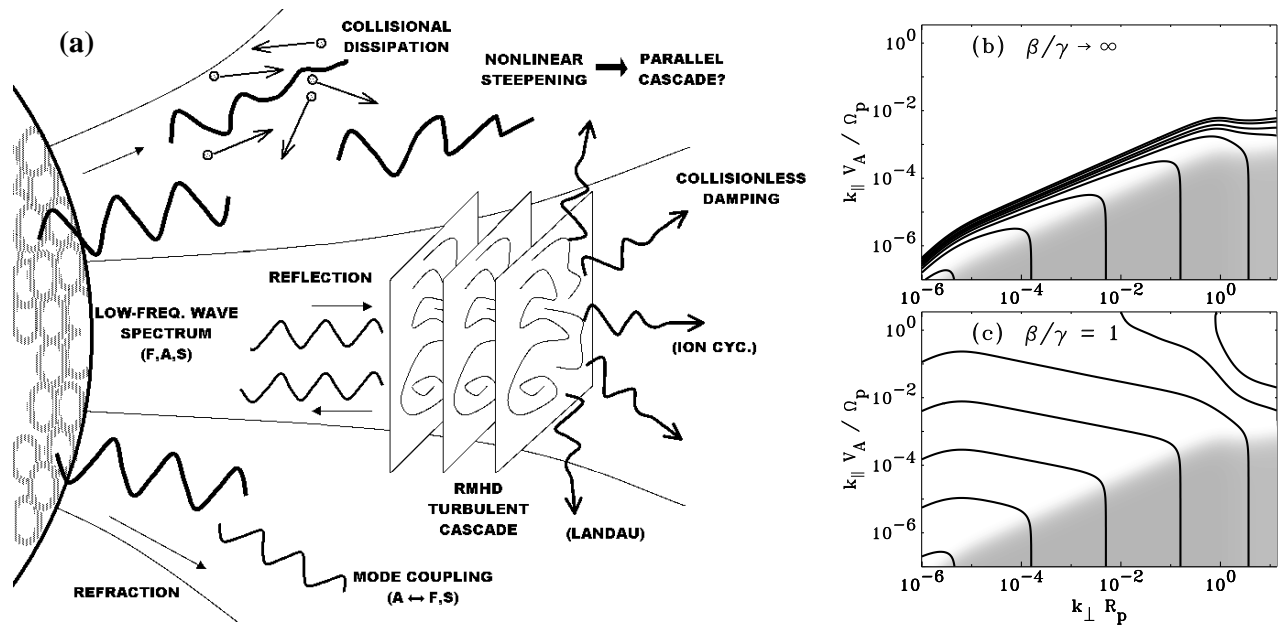

*Figure 6.* (a) *Schematic representation of various physical processes that are believed to be important in heating and accelerating particles in the extended corona (inspired by Fig. 1 of Oughton et al. 2001).* (b, c) *Contours of modeled MHD turbulence power spectra, plotted one per $10^5$ (solid lines) for models with (b) negligible and (c) moderate $k$-space diffusion. $R_p$ denotes the mean proton gyroradius. Gray regions denote the wavenumbers expected to have substantial power given the Goldreich-Sridhar (1997) spectral anisotropy (see Cranmer and van Ballegooijen 2003).*

For more information, see:
http://cfa–www.harvard.edu/∼scranmer/